\documentclass[prl,twocolumn,preprintnumbers,amsmath,amssymb]{revtex4}

\usepackage{bm,curves}
\usepackage{epsfig}
\usepackage[colorlinks=true,linkcolor=blue]{hyperref} 
\usepackage{color}
\usepackage{dcolumn}
\DeclareGraphicsExtensions{.pdf}

\begin{document}

\preprint{JLAB-THY-14-1930}

\title{New limits on intrinsic charm in the nucleon
	from global analysis \\ of parton distributions}

\author{P. Jimenez-Delgado$^1$,
        T. J. Hobbs$^{2,3}$,
        J. T. Londergan$^2$,
        W. Melnitchouk$^1$}

\affiliation{
	$^1$\mbox{Jefferson Lab, 12000 Jefferson Avenue,
         Newport News, Virginia 23606, USA} \\
        $^2$Department of Physics and
         Center for Exploration of Energy and Matter,
         Indiana University, Bloomington, Indiana 47405, USA \\
        $^3$\mbox{Department of Physics,
         University of Washington, Seattle, Washington 98195, USA}}

\date{\today}

\begin{abstract}
We present a new global QCD analysis of parton distribution
functions, allowing for possible intrinsic charm (IC)
contributions in the nucleon inspired by light-front models.
The analysis makes use of the full range of available
high-energy scattering data for $Q^2 \gtrsim 1$~GeV$^2$
and $W^2 \gtrsim 3.5$~GeV$^2$, including fixed-target
proton and deuteron cross sections at lower energies
that were excluded in previous global analyses.
The expanded data set places more stringent constraints
on the momentum carried by IC, with
        $\langle x \rangle_{_{\rm IC}}$ at most 0.5\%
(corresponding to an IC normalization of $\sim 1\%$)
at the 4$\sigma$ level for $\Delta\chi^2 = 1$.
We also critically assess the impact of older EMC measurements
of $F_2^c$ at large $x$, which favor a nonzero IC, but with
very large $\chi^2$ values.
\end{abstract}
\maketitle

%%%%%%%%%%%%%%%%%%%%%%%%%%%%%%%%%%%%%%%%%%%%%%%%%%%%%%%%%%%%%%%%%%%%%%%%
There has been considerable interest recently in the nature of Fock
states of the proton wave function involving five or more quarks,
such as $| uudq\bar q \rangle$, where $q = u, d, s$ or $c$
\cite{Speth98, Kumano98, Vogt00, Garvey01, Chang11, Peng14}.
This has arisen partly from attempts to understand flavor asymmetries
observed in the nucleon sea, such as $\bar d > \bar u$
\cite{NMC, E866} and $s \neq \bar s$ \cite{Mason07}, which clearly
point to a nonperturbative origin.
In addition, there has been a long-standing debate about the existence
of intrinsic charm (IC) quarks in the proton, associated with the
$| uudc\bar c \rangle$ component of the proton wave function.

Aside from the intrinsic interest in the role of nonperturbative
dynamics in the structure of the nucleon sea, the leptoproduction
of charm quarks is also important in providing information on the
gluon distribution in the nucleon.
A significant IC component in the nucleon wave function could
also influence observables measured at the LHC, either directly
through enhanced cross sections at large $x$, or indirectly via
the momentum sum rule leading to a decreased momentum fraction
carried by gluons.

Following early indications from measurements of charm production in
$pp$ scattering of an anomalous excess of $D$ mesons at large values of
Feynman $x_F$ (see \cite{Hobbs14} and references therein), the proposal
was made that the observed enhancement could be accounted for with
the addition of intrinsic $c\bar c$ pairs in the nucleon that were
not generated through perturbative gluon radiation \cite{BHPS}.
Neglecting quark transverse momentum and assuming a charm mass much
greater than other mass scales, Brodsky, Hoyer, Peterson and Sakai
(BHPS) \cite{BHPS} derived an analytic approximation to the IC
distribution that, unlike the perturbatively generated charm,
was peaked at relatively large parton momentum fractions $x$.

A number of experimental and theoretical studies have since
sought to elucidate this issue, although the evidence
has been somewhat inconclusive.
Measurements of the charm structure function $F_2^c$ by the   
European Muon Collaboration (EMC) \cite{EMC} provided
tantalizing evidence for an enhancement at large $x$; however,
more recent experiments at HERA \cite{HERA} at small $x$
found significant tension with the EMC data in regions of
overlapping kinematics.

Early theoretical analyses of the EMC charm data indicated an IC
component with normalization
	$N_{_{\rm IC}} \equiv \int_0^1 dx\, c(x) \sim 1\%$,
although later, more sophisticated treatments incorporating the
photon--gluon fusion (PGF) process, as well as quark and target mass
corrections, argued for smaller IC, $\sim 0.3\%$ \cite{Hoffmann83}.
A subsequent study by Harris, Smith and Vogt \cite{Harris96}
which included ${\cal O}(\alpha_s$) corrections to the hard
scattering cross section obtained a best fit to the highest-energy
EMC data with $N_{_{\rm IC}} = (0.86 \pm 0.60)\%$.
A follow-up analysis by Steffens {\it et al.} \cite{Steffens99}
employed a hybrid scheme to interpolate between massless evolution
at large $Q^2$ and PGF at low $Q^2$, using the BHPS IC model and
a model based on fluctuations of the nucleon to charmed baryon
and $D$ meson states \cite{SaoPaulo, MT97, Pumplin06}.
While it was difficult to fit the data simultaneously in terms of a
single IC framework, Steffens {\it et al.} found a slight preference
for IC in the meson-baryon model at a level of
$N_{_{\rm IC}} \approx 0.4\%$.

To place the study of IC on a more robust statistical footing,
Pumplin {\it et al.} \cite{Pumplin07} used the framework of the
CTEQ global fit \cite{CTEQ6.5} to parton distribution functions
(PDFs) to determine the level of IC that could be accommodated by
the high-energy data.
Comparing the BHPS model, a $p \to \Lambda_c^+ \bar{D}^0$ fluctuation
model with scalar couplings, and a sea-like {\it ansatz} in which
the charm distribution is proportional to the $\bar u$ and $\bar d$
PDFs, the analysis found an allowed range of IC from zero to a
level 2--3 times larger than earlier estimates \cite{Pumplin07}.

An updated NNLO fit by Dulat {\it et al.} \cite{Dulat14}, based on the
more recent CT10 global analysis \cite{CT10} and the BHPS and sea-like
IC models, found the momentum fraction carried by intrinsic charm quarks,
\begin{equation}
\langle x \rangle_{_{\rm IC}}
\equiv \int_0^1 dx \, x \, [c(x) + \bar{c}(x)],
\label{eq:momfrac}
\end{equation}
to be $\lesssim 2.5\%$ for the BHPS distribution at the 90\%
confidence level.  Note that for the BHPS distribution with a 1\%
normalization, the corresponding momentum fraction is
$\langle x \rangle_{_{\rm IC}} = 0.57$\%.
The Dulat {\it et al.} analysis therefore suggests that the existing
data may tolerate rather significant momentum carried by IC.

In this letter, we revisit the question of the magnitude of IC
allowed by the world's $F_2^c$ and other high-energy data,
by performing a new global QCD analysis, along the lines of the
recent JR14 fit \cite{JR14}.
Unlike previous global analyses \cite{Pumplin07, Dulat14} which
placed more stringent cuts on the data ($Q^2 \gtrsim 4$~GeV$^2$
and $W^2 \gtrsim 12$~GeV$^2$), excluding for instance all fixed
target measurements from SLAC \cite{SLAC}, we include all available
data sets with $Q^2 \geq 1$~GeV$^2$ and $W^2 \geq 3.5$~GeV$^2$.
Since most IC models predict this effect to be most prominent
at large values of $x$, excluding the largest-$x$ data may
seriously reduce the sensitivity of the global fit to any IC
that may be present.
In addition, we assess the consistency of the EMC $F_2^c$
data \cite{EMC}, which have often been cited as providing the
strongest evidence for IC in high-energy processes.

Of course, inclusion of lower-$Q^2$ data requires careful treatment of
finite-$Q^2$ and nuclear corrections at intermediate and large $x$.
Following Refs.~\cite{ABKM, CJ11, CJ12, JR14}, we account for
target mass corrections explicitly, using the moment space results
for $F_2$ and $F_L$ from Ref.~\cite{Steffens12}, and allow for
phenomenological $1/Q^2$ higher twist contributions.
For nuclear smearing and nucleon off-shell corrections in the
deuteron, we adopt the method used in the CJ global analysis
\cite{CJ11, CJ12}, while for data on heavier nuclei the nuclear
PDFs from Ref.~\cite{DS04} are employed.
Also, whenever possible, we fit the original cross section data
rather than structure functions derived using an assumed
longitudinal to transverse cross section ratio.
(Further details about the QCD analysis can be found
in Ref.~\cite{JR14}.)

For the QCD analysis we use the framework of the JR14 global fit
\cite{JR14}, in which the $F_2$ structure function
\begin{equation}
F_2 = F_2^{\rm light} + F_2^{\rm heavy}
\label{eq:F2def}
\end{equation}
is decomposed into ($u$, $d$, $s$) quark and heavy ($c, b$)
quark contributions.
The charm structure function is further decomposed into perturbative
($F_2^{c\bar c}$) and nonperturbative ($F_2^{\rm IC}$) components,
\begin{equation}
F_2^c = F_2^{c\bar c} + F_2^{\rm IC}.
\label{eq:F2c} 
\end{equation}
The perturbative part is computed in the fixed-flavor number
scheme (FFNS) from the PGF process \cite{PJD-thesis},
\begin{equation}
F_2^{c\bar c}(x,Q^2,m_c^2)
= \frac{Q^2 \alpha_s}{4\pi^2 m_c^2}
  \sum_i \int\frac{dz}{z}\,
  \hat\sigma_i(\eta,\xi)\, f_i\Big(\frac{x}{z}, \mu\Big),
\label{eq:F2cc}
\end{equation}
where $\hat\sigma_i$ is the hard scattering cross section for
the production of a $c\bar c$ pair from a parton of flavor $i$
($i = u, d, s$ or $g$), and $f_i$ is the corresponding parton
distribution, both calculated to NLO [${\cal O}(\alpha_s)$] accuracy.
The partonic cross section $\hat\sigma_i$ is evaluated as
a function of the scaling variables $\xi = Q^2/m_c^2$ and
$\eta = Q^2(1-z)/(4m_c^2 z) - 1$, and the PDF is computed
at the factorization scale $\mu^2 = 4 m_c^2 + Q^2$.
The analysis therefore fully takes into account the kinematical
corrections arising from quark and target mass effects.
In the FFNS the charm mass does not enter the evolution equations
directly (only indirectly through the running of $\alpha_s$).
For the running mass of the charm quark we take $m_c(m_c) = 1.3$~GeV
at the charm scale in the $\overline{\rm MS}$ scheme; reasonable
variations in the value of $m_c$ have only slight impact on the
results and do not affect our conclusions.

For the nonperturbative charm contributions to $F_2^c$, we consider
several models from recent IC analyses, including variants of the
meson-baryon fluctuation model used to describe charmed baryon
production in hadronic collisions \cite{Hobbs14}, and the BHPS
five-quark model \cite{BHPS}.  The meson-baryon model of
Ref.~\cite{Hobbs14} includes virtual meson-baryon configurations
with pseudoscalar $\bar D$ and vector $\bar{D}^*$ mesons of mass
up to $\approx 2$~GeV, and spin-1/2 ($\Lambda_c$, $\Sigma_c$)
and spin-3/2 ($\Sigma_c^*$) charm baryons.
In contrast, the meson-baryon model of Pumplin \cite{Pumplin06}
generated IC distributions from the fluctuation of the nucleon
to a scalar $\bar D\, \Lambda_c$ state.
To regulate the short-distance behavior of the hadronic loop
integrals, a Gaussian form factor was introduced to dampen the
high-momentum components of the hadronic light-cone distributions,
with the cut-off parameter fit to reproduce the inclusive charmed
baryon and meson production in $NN$ collisions \cite{Hobbs14}.

The IC distributions in the nucleon were then obtained by
convoluting the charmed meson and baryon distributions with
the corresponding PDFs in the charmed hadrons.
%
% For the latter, in addition to a $\delta$-function {\it ansatz}
% motivated by the heavy quark limit \cite{MT97}, two different
% methods were employed in Ref.~\cite{Hobbs14} to regulate
% singularities in the quark propagators: the first used an
% effective charm mass, as in Ref.~\cite{Pumplin06}, while the
% second removed the poles by modifying the propagator in a way
% designed to simulate the effects of confinement \cite{Hobbs14, MST94}.
%
In the present analysis we fix the shapes of the IC distributions
computed in Ref.~\cite{Hobbs14} by the respective best fit cut-off
parameters, but allow the overall normalization to vary.
Provided the variation of the cut-offs is not dramatic,
the effect on the shape of the IC distribution is minor.
Note that the meson-baryon fluctuation model naturally
accommodates asymmetric $c$ and $\bar c$ distributions
as a function of $x$ \cite{MT97}.
Finally, from the IC distribution in a given model, the IC
contribution to the charm structure function is computed using
the framework of Hoffman and Moore \cite{Hoffmann83} to
${\cal O}(\alpha_s)$.

\begin{figure}[t]
\includegraphics[width=8cm]{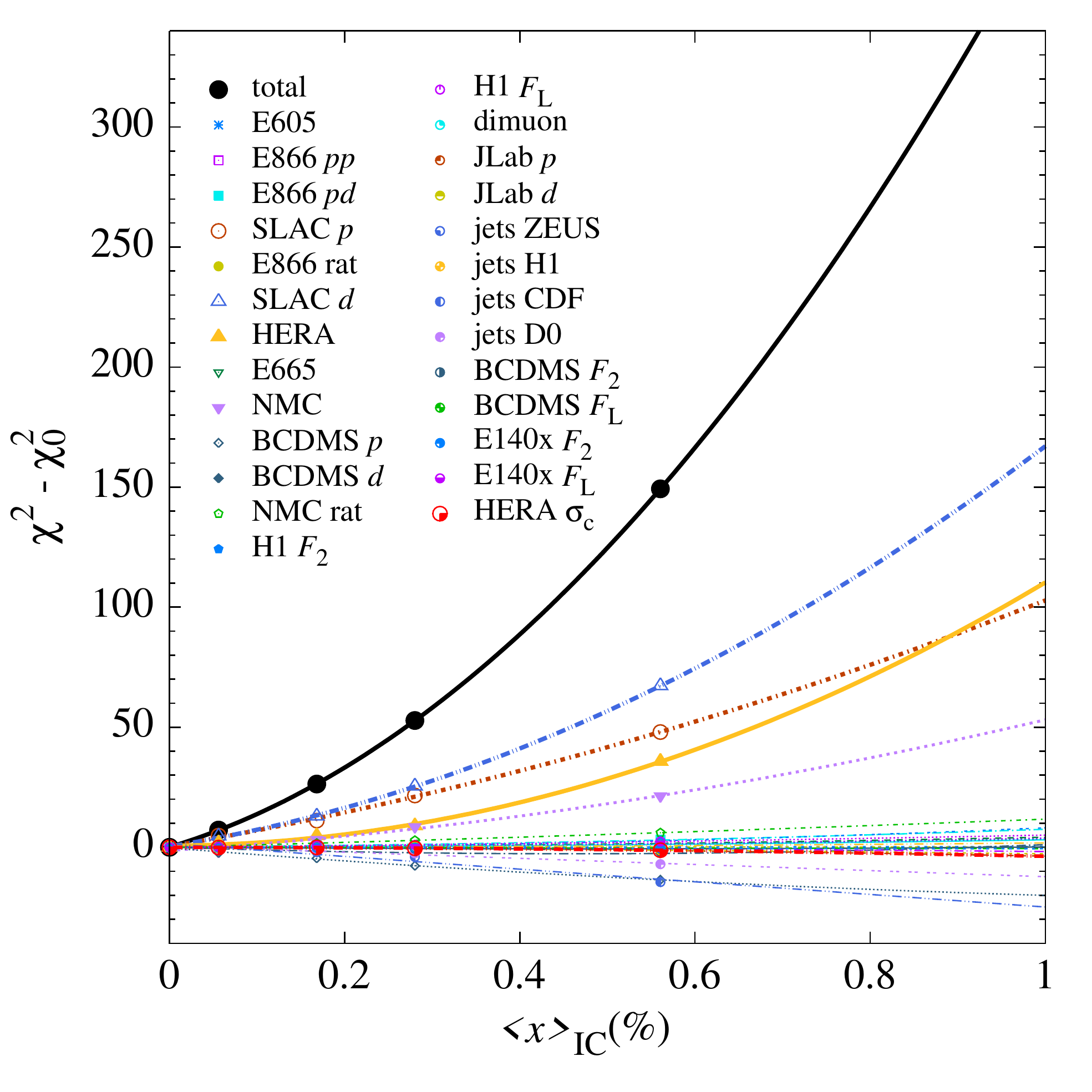}
\caption{(color online)
	Contributions to the total $\chi^2$ (black circles),
	relative to the value $\chi_0^2$ for no IC, of various
	data sets as a function of the momentum fraction
	$\langle x \rangle_{_{\rm IC}}$ carried by IC quarks
	(in percent).
	The largest contributions to the total $\chi^2$ are from
	the SLAC inclusive deuteron (blue triangles) and proton
	(brown circles) structure functions, HERA $F_2^c$
	(orange triangles) and NMC $F_2$ (violet triangles) data.
	The EMC $F_2^c$ data are not included in this fit.}
\label{fig:scan_full}
\end{figure}

The results of the global analysis are summarized in
Fig.~\ref{fig:scan_full}, where the total $\chi^2$ values for
each of the 26 data sets used in the fit are shown (relative to
the value for no IC, $\chi_0^2$) as a function of the momentum
fraction carried by IC quarks.
The total $\chi^2$ has its minimum for zero IC, and rises
rapidly with increasing $\langle x \rangle_{_{\rm IC}}$.
The largest contributions to $\chi^2$ arise from the SLAC
deep-inelastic proton and deuteron structure functions \cite{SLAC},
with smaller contributions from HERA charm production at low $x$
\cite{HERA}, and NMC proton and deuteron cross sections in the
medium-$x$ region \cite{NMC-data}.
All other data sets have little or no sensitivity to IC,
as evidenced by the rather shallow $\chi^2$ profiles.
The total $\chi^2$ for the global fit gives
$\chi^2/N_{\rm dat} = 1.25$ for $N_{\rm dat} = 4296$ data points.

\begin{figure}[t]
\includegraphics[width=8cm]{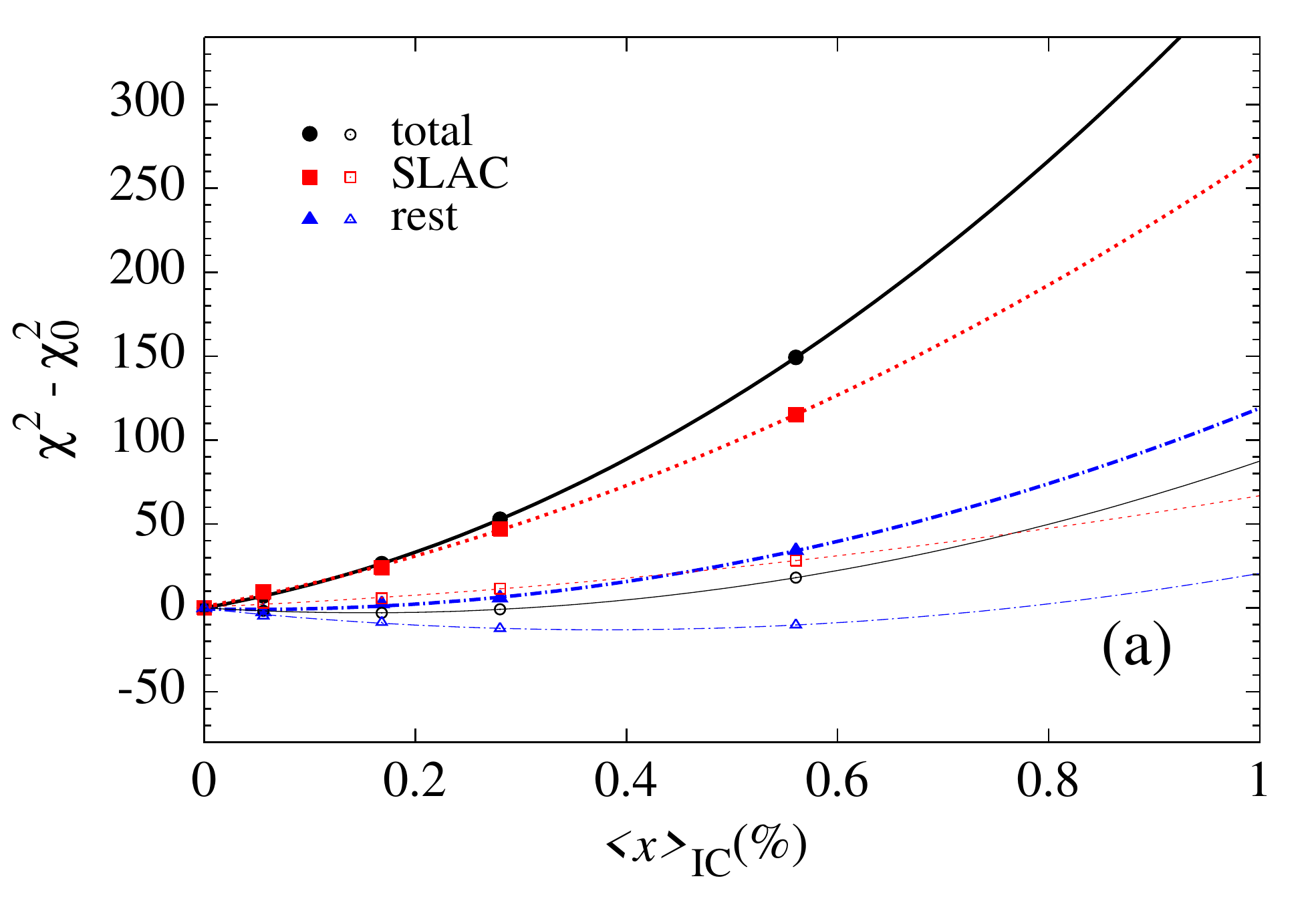}
\includegraphics[width=8cm]{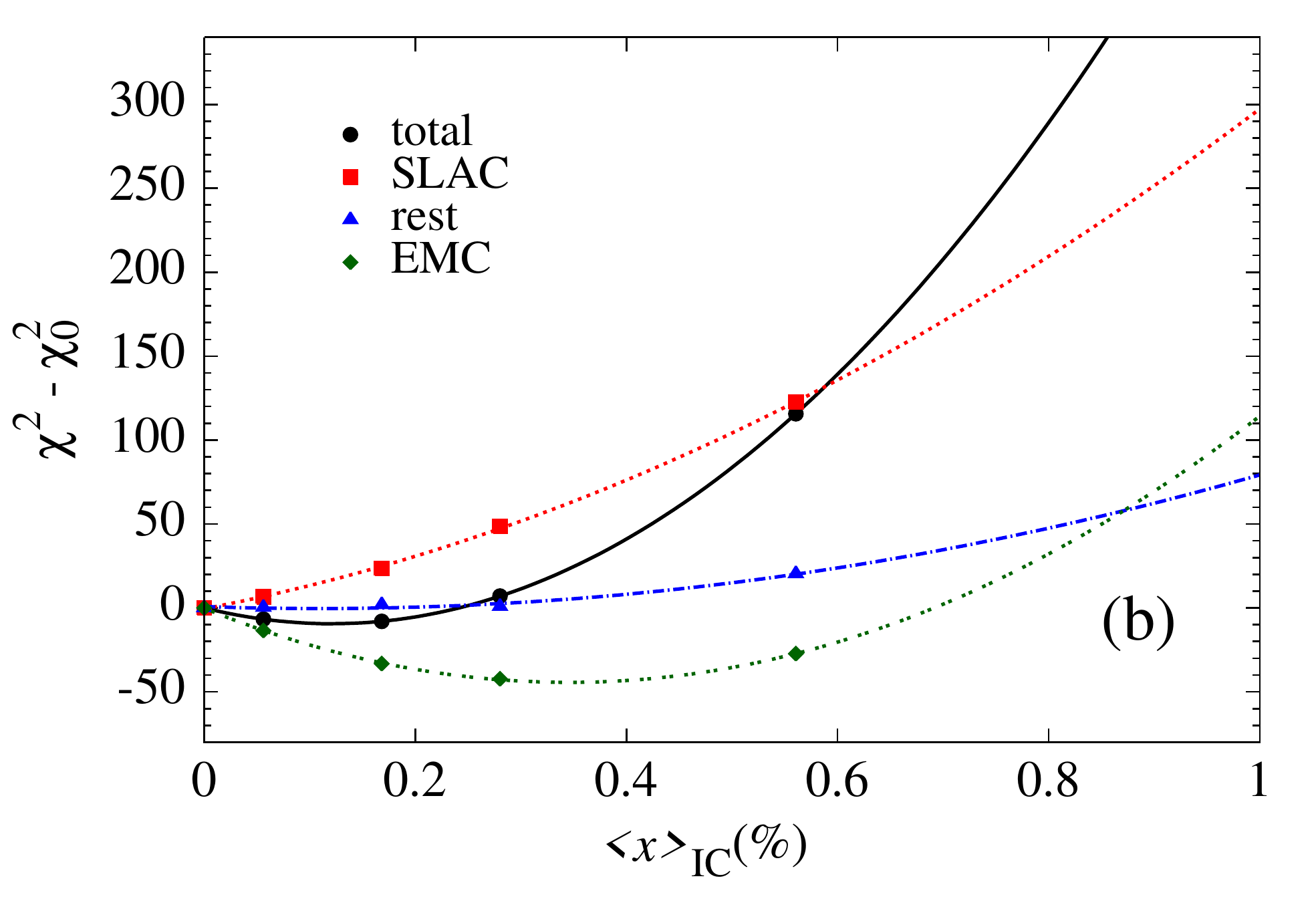}
\caption{(color online)
	Contributions of various data sets to the total
	$\chi^2$, relative to $\chi_0^2$, as a function	of
	$\langle x \rangle_{_{\rm IC}}$ (in percent) for
	(a) the standard data set, and
	(b) including the EMC $F_2^c$ data.
	In (a), the upper curves (filled symbols) represent
	the standard fit, while the lower curves (open symbols)
	include a threshold suppression factor \cite{Brodsky14}.}
\label{fig:scans}
\end{figure}

Because of the more restrictive $Q^2$ and $W^2$ cuts employed in
previous global IC studies \cite{Pumplin07, Dulat14}, which were
tuned more to collider data, lower energy fixed-target data
such as from SLAC were excluded from the fits.
This produced rather weak limits on the IC momentum fraction,
$\langle x \rangle_{_{\rm IC}} \lesssim 2-3\%$.
Including the full data set, we find a much more stringent
constraint on the momentum carried by IC, with
	$\langle x \rangle_{_{\rm IC}} < 0.1\%$
at the 5$\sigma$ level.
The rest of the $\chi^2$ profile allows slightly larger IC
values, as illustrated in Fig.~\ref{fig:scans}(a), with
	$\langle x \rangle_{_{\rm IC}} < 0.1\%$
at the 1$\sigma$ level.
%
% Higher momentum fraction limits of $0.2\%$ are excluded at the
% 2$\sigma$ level, and $0.3\%$ excluded at the 3$\sigma$ level.

Note that a significant portion of the SLAC data (360 points
from a total of 1021) lie below the partonic charm threshold,
$W^2 < 4 m_c^2$,
so that these data do not provide direct constraints on IC.
However, through $Q^2$ evolution the stronger constraints
on the light-quark PDFs at \mbox{high $x$} from the low-$W$ region
allow important limitations on the magnitude of the IC to be
obtained from the global fit to the expanded data set.
In fact, the partonic threshold is lower than the physical
threshold at which charmed hadrons can be produced, which in DIS
would correspond to $W^2 > W_{\rm thr}^2 \approx 16$~GeV$^2$.
Even above this value there are still 157 data points in the
SLAC $p$ and $d$ data sets.

To take into account the mismatch between the partonic and hadronic
charm thresholds, various prescriptions have been adopted in the
literature.  The MSTW analysis \cite{MSTW} employed a ``modified
threshold'' approach with an effective charm quark mass
$m_c (1+\Lambda^2/m_c^2)$ in the threshold dependent parts of
coefficient functions, where $\Lambda$ is a ``binding energy''
parameter.  An alternative prescription \cite{Brodsky14} advocates
a phase space factor
	$\theta(W^2-W_{\rm thr}^2) (1-W_{\rm thr}^2/W^2)$
weighting $F_2^c$ in Eq.~(\ref{eq:F2c}) to suppress charm
contributions near threshold.
The fits with the hadron suppression factor, illustrated in
Fig.~\ref{fig:scans}(a), show a generally shallower $\chi^2$
profile, with
  $\langle x \rangle_{_{\rm IC}}$ at most $\approx 0.5\%$
at the 4$\sigma$ level.
The minimum $\chi^2$ in this case occurs at
  $\langle x \rangle_{_{\rm IC}} = (0.15 \pm 0.09)\%$
for the full \mbox{data set}.

The differences between our analysis without the SLAC data and those
in Refs.~\cite{Pumplin07, Dulat14} are partly explained by the
different tolerance criteria used: in our fits the PDF errors refer
to variations of $\Delta\chi^2 = 1$ around the minimum, whereas the
previous analyses \cite{Pumplin07, Dulat14} assumed a tolerance of
$\Delta\chi^2 = 100$.  There is no unique criterion for selecting
the correct $\Delta\chi^2$ interval, and we use the traditional
$\Delta\chi^2 = 1$ choice based on statistical considerations alone.
Choosing $\Delta\chi^2 = 100$ would inflate the uncertainty and
accommodate $\langle x \rangle_{_{\rm IC}} \approx 1\%$ at the
1$\sigma$ level, which is comparable to that in the earlier work.

While the global fits in Fig.~\ref{fig:scan_full} incorporate the
charm production cross sections from HERA \cite{HERA}, they do not
include the earlier charm structure function data from EMC \cite{EMC}.
Since the HERA cross sections are predominantly measured at small $x$,
they have less sensitivity to the presence of IC than the fixed-target
data at larger $x$, as the $\chi^2$ profile in Fig.~\ref{fig:scan_full}
illustrates.
On the other hand, the EMC $F_2^c$ measurements include data at
large $x$ values, which have greater impact on the IC determination.
In Fig.~\ref{fig:scans}(b) the $\chi^2$ values for the global fits
including the EMC data indicate a slight preference for a nonzero IC,
with the EMC data alone favoring a value $\sim (0.3-0.4)\%$
(the additional threshold suppression factor has a minor impact
on the EMC data).
However, the description of the EMC data is clearly far from
satisfactory, giving a $\chi^2$ value of 4.3 per datum for
19 data points.

\begin{figure}[t]
\includegraphics[width=8.5cm]{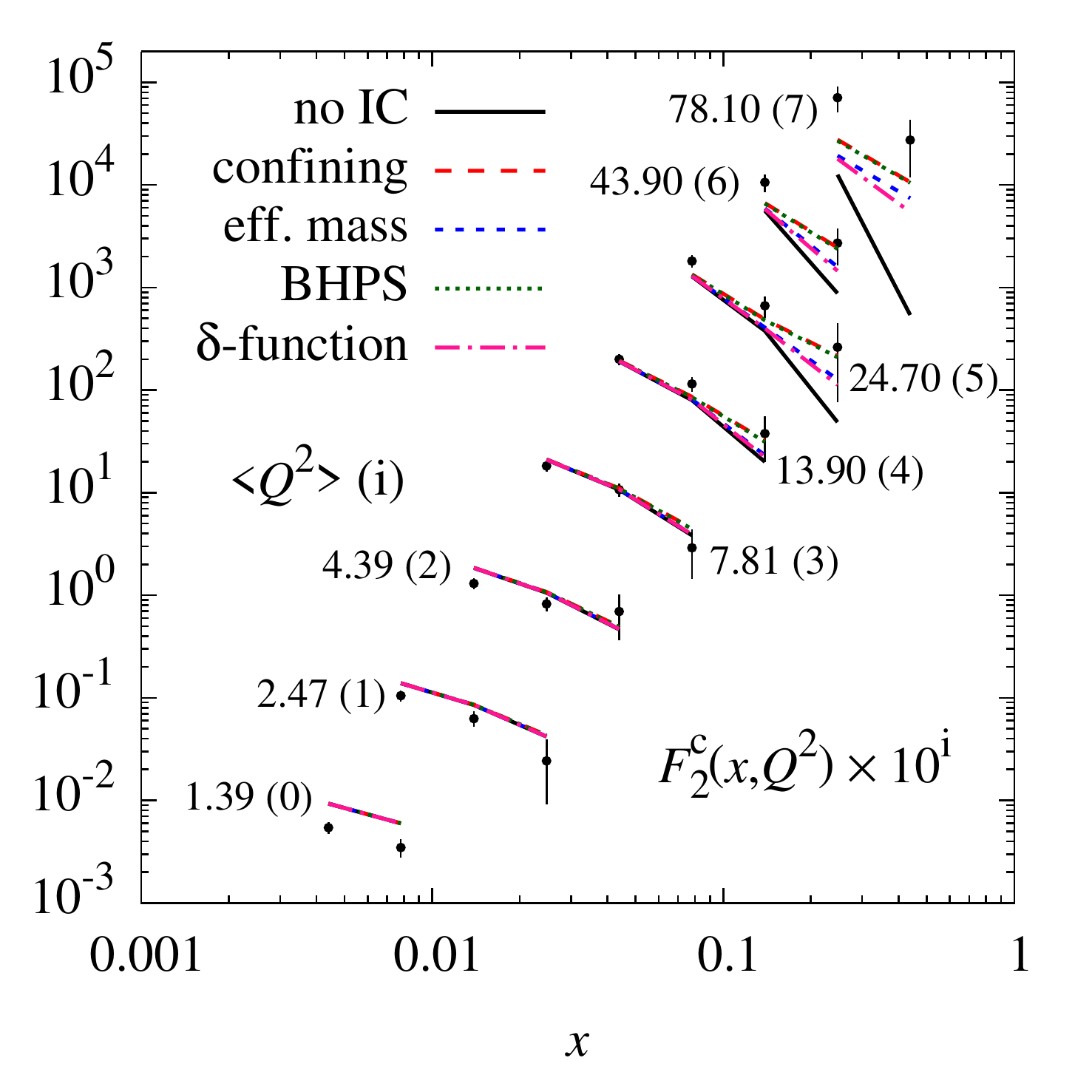}
\caption{(color online)
	Comparison of the total fitted $F_2^c$ structure function
	with the full set of EMC data \cite{EMC} for $Q^2$ between
	1.39~GeV$^2$ to 78.1~GeV$^2$.
	The results with no IC (black solid lines) are compared
	with those using the confining (red dashed lines),
	effective mass (blue short-dashed lines),
	BHPS (green dotted lines), and
	$\delta$-function (pink dot-dashed lines)
	models for IC \cite{Hobbs14}.}
\label{fig:F2cEMC}
\end{figure}

The comparison with the full set of $F_2^c$ data from EMC is
shown in Fig.~\ref{fig:F2cEMC} for several models of IC from
Refs.~\cite{BHPS, Hobbs14}, as well as for a fit without IC.
At small $x$ values ($x \lesssim 0.02$) the global fits
generally overestimate the data, regardless of whether IC
(which is negligible in this region) is included or not.
At intermediate $x$ ($0.02 \lesssim x \lesssim 0.1$), where
the IC contributions are still small, the agreement improves,
while at the largest $x$ values ($x \gtrsim 0.2$) the fit
with no IC clearly lies below the data.
Here the addition of IC improves the agreement for all models
considered, with the meson-baryon model for the confining
$c$ quark-diquark interaction \cite{Hobbs14} and the BHPS
model \cite{BHPS} resulting in the biggest enhancement.
On the other hand, the experimental uncertainties at the high
$x$ values are rather large compared with those in the small-$x$
region, where the fit to the EMC $F_2^c$ data is worse.
Better agreement with the EMC data would require significantly
larger IC at high $x$, together with some additional suppression
mechanism at low $x$ values, neither of which appear very probable.
Because of the significant tension with the other global data sets,
the EMC data are usually not included in most global PDF analyses
\cite{CTEQ6.5, Dulat14, CT10, JR14, ABKM, CJ11, CJ12, MSTW}.

These conclusions are more consistent with those reached in the
MSTW analysis \cite{MSTW}, which found reasonable fits including
the EMC data for $N_{_{\rm IC}} = 0.3\%$ using the BHPS model.
On the other hand, the analysis \cite{MSTW} also utilized more
stringent cuts ($Q^2 \geq 2$~GeV$^2$ and $W^2 \geq 15$~GeV$^2$)
than those used in our fit, which removed much of the SLAC data
at large $x$, and did not consider higher twist corrections ---
both of which are important in the region where IC is expected
to contribute.

In summary, we have performed a comprehensive global QCD analysis
of the world's high-energy scattering data, synthesizing the latest
developments in global fitting technology and nonperturbative studies
of charm production to fully exploit all of the available data
that may have bearing on the question of IC in the nucleon.
By relaxing the cuts on $Q^2$ and $W^2$ used in earlier global fits
\cite{MSTW, Pumplin07, Dulat14}, while systematically accounting
for finite-$Q^2$ and other hadronic and nuclear corrections
\cite{ABKM, CJ12, JR14}, we found that the low-$Q^2$, high-$x$
data from fixed-target experiments in particular place stronger
constraints on the magnitude of IC than found previously.
Excluding the older $F_2^c$ measurements from the EMC \cite{EMC},
which give a very large $\chi^2$, our fits generally rule out large
values of IC, with $\langle x \rangle_{_{\rm IC}}$ at most 0.5\%
at the 4$\sigma$ level, even after taking into account
nonperturbative charm threshold suppression factors.
The tension between the EMC data and the more precise measurements
of $F_2^c$ at HERA at low $x$ \cite{HERA} has prompted many global
PDFs analyses to omit these data from their fits.
Given that the signal for IC relies so heavily on charm production
data at large values of $x$, it would be essential to obtain new,
more precise data on $F_2^c$ to determine limits (upper or lower)
on the nonperturbative charm content of the nucleon with greater
confidence.  Such measurements could be feasible at a future
electron-ion collider facility \cite{EIC}.

%%%%%%%%%%%%%%%%%%%%%%%%%%%%%%%%%%%%%%%%%%%%%%%%%%%%%%%%%%%%%%%%%%%%%%%%
T.J.H. and J.T.L. were supported in part by the U.S. National 
Science Foundation under Grant No.~NSF-PHY-1205019.
The work of T.J.H. was also supported in part by DOE Grant
No.~DE-FG02-87ER40365. W.M. was supported by the DOE Contract
No.~DE-AC05-06OR23177, under which Jefferson Science Associates,
LLC operates Jefferson Lab.

%%%%%%%%%%%%%%%%%%%%%%%%%%%%%%%%%%%%%%%%%%%%%%%%%%%%%%%%%%%%%%%%%%%%%%%%%

\end{document}